\documentclass[prc,twocolumn,floatfix,groupedaddress,nofootinbib,showpacs,preprintnumbers,
amsmath,amssymb,amsfonts,superscriptaddress,widetable] {revtex4}
\usepackage{bm}
\usepackage{mathrsfs}
\usepackage{amssymb}
\usepackage{amsmath}
\usepackage{graphicx}
\usepackage{array}
\usepackage{color}

\begin{document}
\title{Crust breaking and the limiting rotational frequency of neutron stars}
\author{F.~J. Fattoyev}\email{ffattoye@indiana.edu}
\affiliation{Center for Exploration of Energy and Matter and
                  Department of Physics, Indiana University,
                  Bloomington, IN 47405, USA}
\author{C. J. Horowitz}\email{horowit@indiana.edu}
\affiliation{Center for Exploration of Energy and Matter and
                  Department of Physics, Indiana University,
                  Bloomington, IN 47405, USA}
\author{Hao Lu}\email{hl64@iu.edu}
\affiliation{Department of Astronomy, Indiana University,
                  Bloomington, IN 47405, USA}
\date{\today}
\begin{abstract}
The limiting rotational frequency of neutron stars may be determined
by the strength of their crusts.  As a star spins up from accretion,
centrifugal forces will cause the crust to fail.  If the crust
breaks unevenly, a rotating mass quadrupole moment will radiate
gravitational waves (GW).   This radiation can prevent further spin
up and may be a promising source for continuous GW searches.   We
calculate that for a breaking strain (strength) of neutron star
crust that is consistent with molecular dynamics simulations, the
crust may fail at rotational frequencies in agreement with
observations.

\end{abstract}
\smallskip

\maketitle

Some neutron stars (NS's) are known to rotate
rapidly\,\cite{Hessels:2006ze}. Initially slowly rotating NS's in
accreting systems can increase their rotational rate as material
falling from a binary companion transfers angular momentum. However,
most of the observed rapidly rotating NS's spin only at about half
of the Keplerian break-up frequency. They are thought to have
reached a spin equilibrium, where the angular momentum gained from
accretion is balanced out by the angular momentum radiated via
either gravitational waves (GW's) or magnetic dipole radiation (See
Ref.\,\cite{Patruno:2017oum} and references therein). In this
Letter, we argue that the limit on the spin frequency of NS's could
be set by the strength of the NS crust.

In the last few decades of the 20th century many works have been
devoted to crust breaking which was explored as a possible source of
starquakes and pulsar glitches\,\cite{Baym:1971xx, Ruderman:1976,
Ruderman:1991c}.  As the NS in the accreted binary system spins
up---or spins down in the case of isolated pulsars---its equilibrium
shape changes and stresses develop in the crust\,\cite{Link:1998,
Franco:2000}.  A sufficiently large stress can break the crust and
change the moment of inertia of the star causing glitches in its
rotation rate. It was recently found, however, that the crust is
likely to be very strong, with a breaking strain (fractional
deformation) of  $\sigma \simeq
0.1$\,\cite{Horowitz:2009ya,Chugunov2010}. The crust is strong
because high pressure suppresses the formation of voids and
fractures and because long range coulomb interactions insure that
each ion is ``bound'' to many other ions.  The breaking strain is
much larger than previously thought
$\sigma\,=\,10^{-4}-10^{-2}$\,\cite{Cutler:2002np}. The failure of a
weak crust could explain some (but not all) of the sudden jumps in
the rotational frequencies of pulsars.  However, a strong crust
allows the star to sufficiently deform before it may undergo
failure. To the best of our knowledge, no previous work has examined
the change in rotational frequency that is necessary for a strong
crust to fail.

A strong crust may allow a star to be spun up considerably before
the crust fails. When a region of the crust fails, it may move under
centrifugal force to larger radii and produce a significant
ellipticity $\epsilon = \left(I_{xx} - I_{yy}\right)/I_{zz}$.  This
is the fractional difference in the principle moments of inertia.  A
strong crust can support an ellipticity as large as $\epsilon \simeq
10^{-5}$\,\cite{JohnsonMcDaniel:2012wg,Horowitz:2009ya}.

When the crust starts to break, we do not know how much of the crust
will fail. If even a small fraction of the crust fails, it could
produce an $\epsilon \sim 10^{-8}$. This $\epsilon$ can radiate GW
and lead to torque balance where the angular momentum gained from
accretion is radiated as GW\,\cite{Patruno:2017oum}. This will
prevent the star from spinning up further.

In this Letter we suggest that {\it the strength of the crust
determines the limiting rotational frequency of neutron stars}. A
star can spin up to a frequency where the crust first starts to
fail.  When the crust does break, it is likely to break unevenly,
for example only in a small region on one side of the star.  This
produces a nonzero $\epsilon$ so that the star radiates GW and is
prevented from spinning up further.  Below we calculate that for
$\sigma$ of order 0.1, the crust may first break at a limiting
rotational frequency of 300-700 Hz.  This is consistent with
observations.  Note that our crust breaking mechanism is distinct
from Bildsten's idea of deformations from nonuniform electron
captures\,\cite{Bildsten1998, Ushomirsky:2000ax} and does not
require a temperature gradient.


There have been many searches for continuous GW from rotating
stars\,\cite{Riles:2017evm}, including from known pulsars, see for
example\,\cite{Abbott:2017cvf}, and from accreting neutron
stars\,\cite{Zhu:2016ghk, Mukherjee:2017qme}.  These searches will
be extended in the near future with Advanced LIGO and VIRGO
data\,\cite{Abbott:2017mnu, Abbott:2017cvf} or by the use of third
generation GW detectors\,\cite{Evans:2016mbw}.

We follow the same formalism of quaking neutron
stars as outlined in Ref.\,\cite{Franco:2000}. The star is modeled
as a two-component homogeneous spheroid: a crust of uniform
density that rests on top of a core of an incompressible fluid. As
the star spins up (or down) a solid crust attempts to re-adjust its
shape, which in turn builds stresses in the crust. The local
distortion of the solid is given by the strain
tensor\,\cite{Landau:1959}:
\begin{equation}
 u_{ij} = \frac{1}{2}\left(\frac{\partial u_i}{\partial x_j} + \frac{\partial u_j}{\partial x_i} \right) \;,
 \label{strain1}
\end{equation}
where $\textbf{u}(\textbf{r})$ is the displacement field in a
material. The strain tensor can be diagonalized in a local
coordinate system, whose positive eigenvalues represent compression
whereas negative eigenvalues represent dilation along the respective
axes. One can then define the strain angle as the difference between
the largest and the smallest eigenvalues $\lambda_{ii}$,
\textit{i.e.}
\begin{equation}
\label{strainangle} \varepsilon \equiv \lambda_{\rm max} -
\lambda_{\rm min} \ .
\end{equation}
The NS crust breaks when the local strain angle in the stress plane
reaches a critical value set by the breaking strain of the material, $\varepsilon=
\sigma$.

The accumulation of strain in spin-down (-up) neutron stars is
described in detail in Ref.\,\cite{Franco:2000}. Here we will
outline the main equations needed to perform our calculations. We
assume that initially the material is under great internal pressure
but is unstrained. For self-gravitating objects, the equation for
hydrostatic equilibrium of a rotating star depends on the material's
stress tensor, the local gravitational potential, and the
centrifugal force. As the star spins slower (faster) the centrifugal
force changes. The condition for hydrostatic equilibrium for this
new configuration is then derived by considering a Lagrangian
perturbation of the initial state. It is shown that the displacement
field in the crust then can be solved using these
expressions\,\cite{Franco:2000}:
\begin{eqnarray}
\label{displacement1} && u_r(r, \theta) = \left(ar -\frac{A}{7}r^3 -
\frac{B}{2r^2} +
\frac{b}{r^4} \right) P_2(\theta) \ , \\
\label{displacement2}  && u_{\theta}(r, \theta) = \left(\frac{ar}{2}
-\frac{5}{42}Ar^3 - \frac{b}{3r^4} \right)
\frac{dP_2(\theta)}{d\theta} \ ,
\end{eqnarray}
where $r$ is the radial distance, $P_2(\theta) = (3 \cos^2 \theta -
1)/2$ is the second Legendre polynomial, $a$, $b$, $A$, and $B$ are
constants that are determined using the following boundary
conditions at the outer and inner boundaries of the crust:
\begin{eqnarray}
\label{coefficients_a} a - \frac{8}{21}AR^2 - \frac{B}{2R^3} +
\frac{8}{3}\frac{b}{R^5} &=& 0  \ , \\
\label{coefficients_b} a - \frac{8}{21}AR^{\prime \,2} -
\frac{B}{2R^{\prime \,3}} +
\frac{8}{3}\frac{b}{R^{\prime \,5}} &=& 0 \ , \\
\label{coefficients_c} -2f^{\prime}(R) -\frac{2}{5} \frac{v_{\rm
K}^2}{c_{\rm t}^2} \frac{f(R)}{R}+ \frac{R^2}{3} \frac{\Omega_{\rm
i}^2 - \Omega_{\rm f}^2}{c_{\rm
t}^2} &=& AR^2+\frac{B}{R^3}, \ \ \ \ \\
\label{coefficients_d} -\frac{1}{2}\left[AR^{\prime \,
2}+\frac{B}{R^{\prime \,3}}\right] &=& f^{\prime}(R^{\prime})  \ .
\end{eqnarray}
Here $R$ and $R^{\prime}$ are the total and core radii of the
neutron star, respectively, $v_{\rm K} \equiv \sqrt{GM/R}$ is the
Keplerian velocity, $c_{\rm t} =\sqrt{\mu/\rho}$ is the transverse
sound speed, $\Omega_{\rm i}$ and $\Omega_{\rm f}$ are the initial
and final angular velocities, and $f(r)$ is the radial part of the
displacement field $u_r(r, \theta)$. The transverse sound speed
depends on the elastic shear modulus $\mu$ \cite{Horowitz:2008xr}
and the density $\rho$. Our calculations show that $c_{\rm t} \simeq
10^{6}$ m/s in most of the crust.

Solving the system of equations consisted of boundary conditions
(\ref{coefficients_a})-(\ref{coefficients_d}) we obtain the four
coefficients ($a$, $A$, $b$, $B$) that go into the equations for the
displacement field, \textit{i.e} Eqns. (\ref{displacement1}) and
(\ref{displacement2}). The result is then applied to the equation
for the strain tensor (\ref{strain1}). One can then find a
coordinate system in which the strain tensor is diagonal, which
gives the strain angle $\varepsilon$ as described in
Eq.\,(\ref{strainangle}).

Instead of fixing the NS radius and crust
thickness\,\cite{Franco:2000}, here we will consider some realistic
cases by choosing several possible configurations stemming from
current uncertainties in the equation of state (EOS) of neutron-star
matter. The size of both the core and the crust is therefore
determined by the EOS. In particular, we select the soft and the
stiff EOS models by Hebeler \textit{et al.} (HLPSsoft and HLPSstiff,
respectively)\,\cite{Hebeler:2010jx}. These two models should
bracket current uncertainties in existing models of the nuclear EOS.
The HLPSsoft predicts a compact star with $R_{1.4} = 9.94$ km,
whereas HLPSstiff predicts a large star with $R_{1.4} = 13.59$ km,
where the index $1.4$ refers to a canonical $1.4 M_{\odot}$ neutron
star.

The size of the crust is crucial in our investigation. Although the
core-crust transition pressures in both models are the same, $P_{\rm
t} = 0.4054$ MeV fm$^{-3}$, the corresponding crustal sizes differ
because of the pressure gradient at the crust-core interface.
Compact stars predicted by a soft EOS have larger pressure gradients
at the crust-core interface due to their strong local gravity and
therefore produce a thinner crust. Whereas the crust of large stars
becomes thick as a result of the weaker gravity at the crust-core
boundary.  Alternatively, one can get a thick crust by using a
moderate EOS that is not relatively stiff, but predicts a crust-core
transition pressure that is large. For this reason, we also select
two relativistic mean-field models, IU-FSU and
IU-FSUmax\,\cite{Fattoyev:2010mx, Piekarewicz:2014lba} which predict
the same maximum stellar masses, but substantially different
crust-core transition pressures, hence crust thicknesses. In
particular, the IU-FSU predicts $P_{\rm t} = 0.2890$ MeV fm$^{-3}$
and $R_{1.4}=12.49$ km, whereas the IU-FSUmax predicts $P_{\rm t} =
0.5184$ MeV fm$^{-3}$ and $R_{1.4}=12.80$ km.


\begin{figure}[ht]
\smallskip
 \includegraphics[width=1.0\columnwidth]{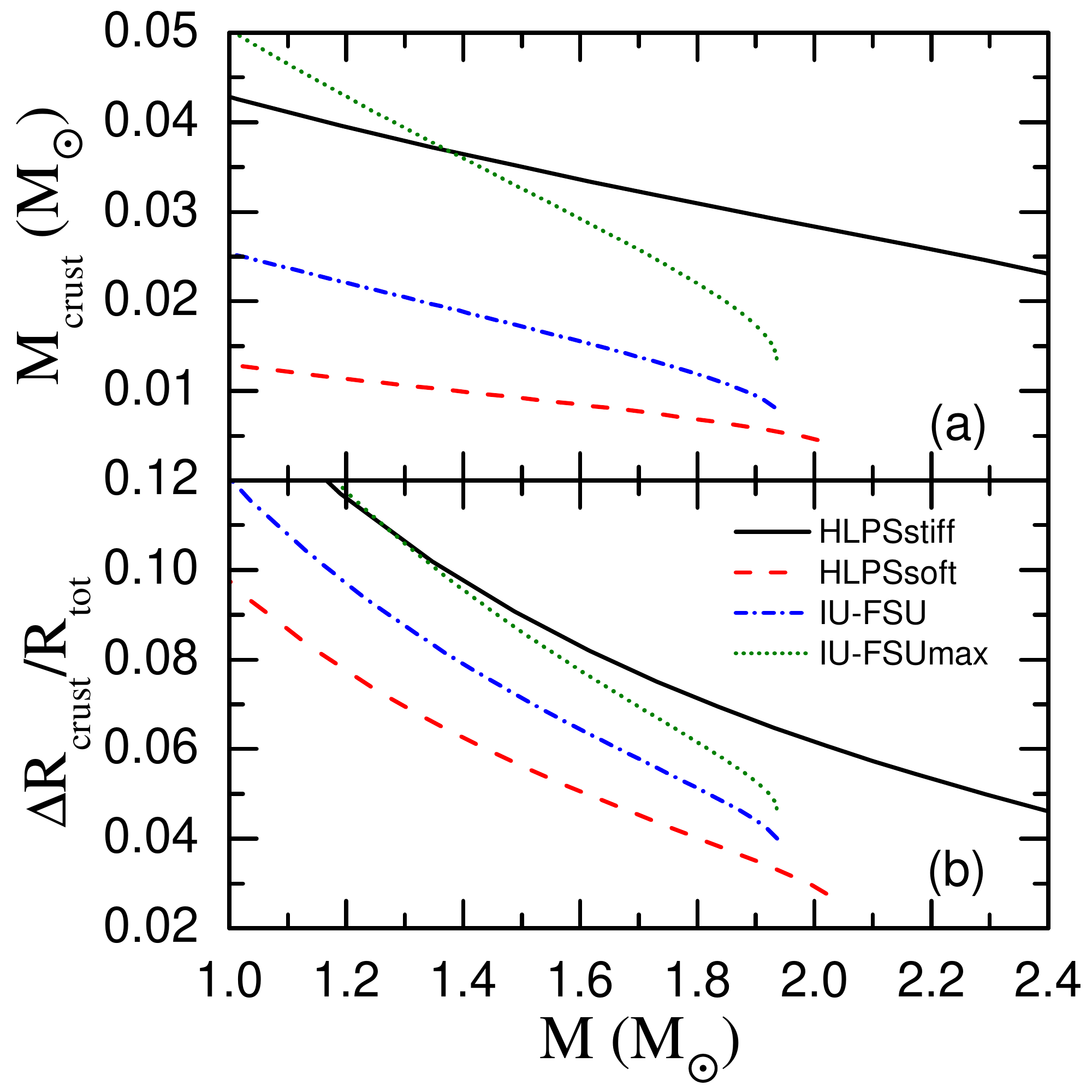}
 \caption{(Color online). The neutron star crust mass $M_{\rm crust}$ (a)
 and crust thickness fraction $\Delta R_{\rm crust}/R_{\rm tot}$ (b) as a function of
 the neutron star mass $M$.}
 \label{Fig1}
\end{figure}

We note that the crust thickness is not only important for our
problem, but also is important in modeling of the cooling of
quiescent low-mass X-ray binaries, as the cooling timescale is
proportional to the square of the crust
thickness\,\cite{Brown:2009kw}. In Fig.\,\ref{Fig1} we plot the
crustal mass $M_{\rm crust}$ (a) and the fraction of crustal
thickness $\Delta R_{\rm crust}/R_{\rm tot}$ (b) as a function of
the total mass of neutron stars predicted by the four models. As
evident from the figure, the crustal mass can be as low as about
$0.01 M_{\odot}$ for compact stars and as large as $0.04 M_{\odot}$
for stars that have a thick crust.

\begin{figure}[ht]
\smallskip
 \includegraphics[width=1.0\columnwidth]{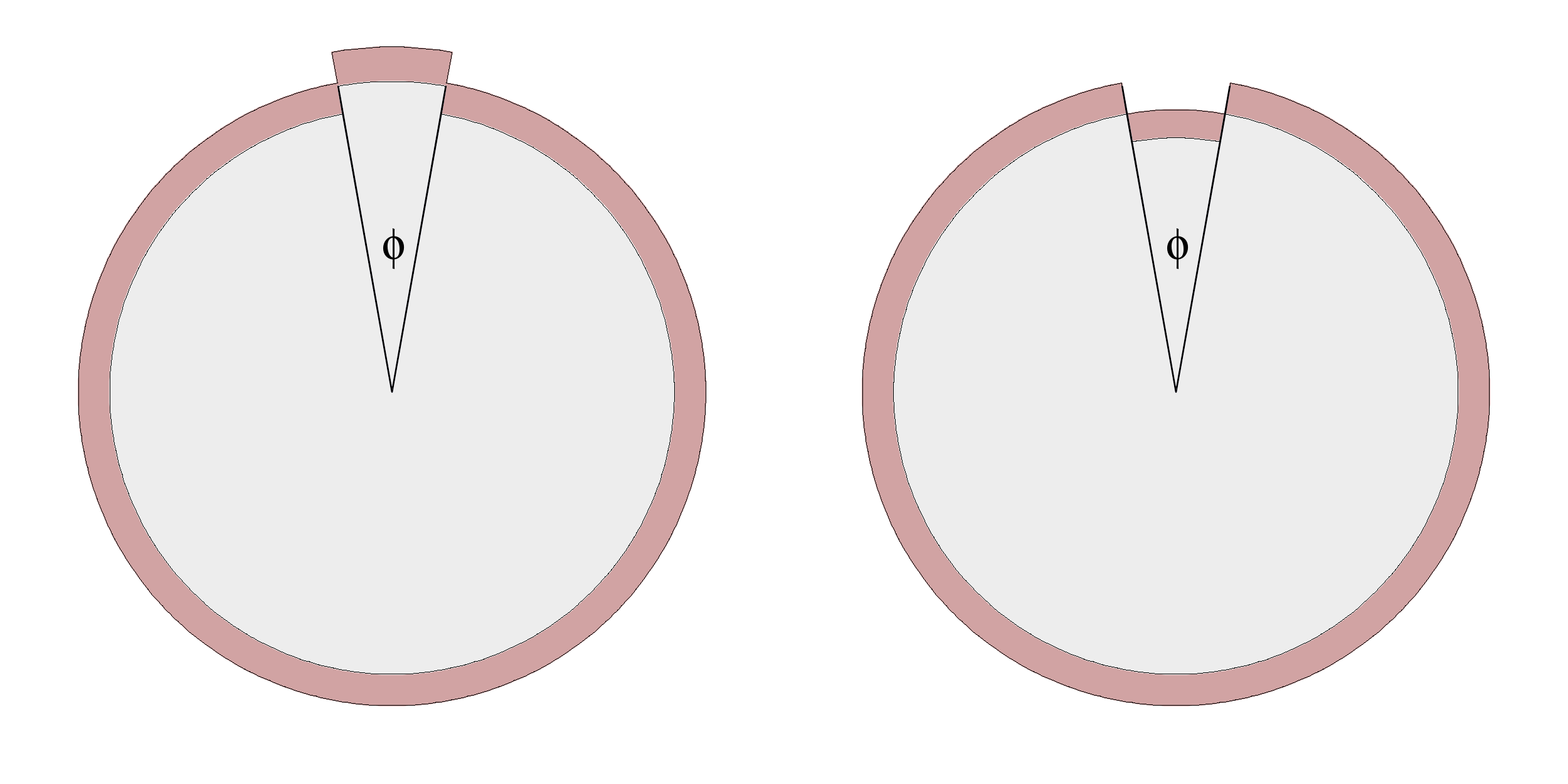}
 \caption{(Color online). (Left) Crust failure on accreting neutron star in arbitrary region $\phi$.
 (Right) Crust failure on isolated slowing down star.  Motion in radial direction under centrifugal force is exaggerated.}
 \label{Fig1b}
\end{figure}

\begin{figure}[ht]
\smallskip
 \includegraphics[width=1.0\columnwidth]{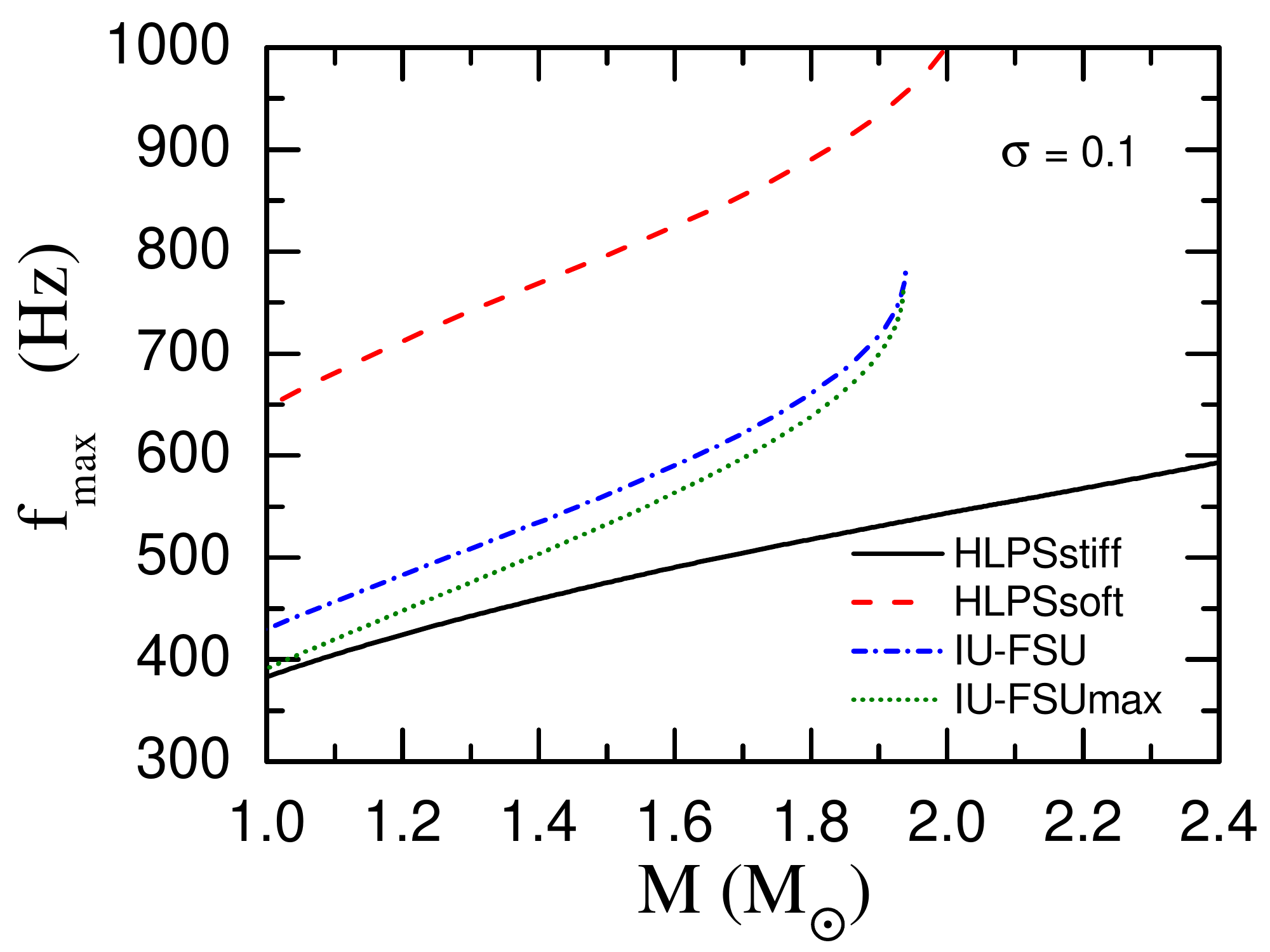}
 \caption{(Color online). The maximum initial rotational frequency of a neutron star
 whose crust will never fail from slowing down, assuming a breaking strain of $\sigma=0.1$.
}
 \label{Fig2}
\end{figure}

\begin{table}[h]
\begin{tabular}{|l||c|c|c|c|}
 \hline
Model                  & $f_{\rm max}^{1.4}$ (Hz)             & $f_{\rm max}^{1.6}$   (Hz)        & $f_{\rm max}^{1.8}$  (Hz) & $f_{\rm max}^{2.0}$  (Hz) \\
 \hline
 \hline
 {\rm HLPSSoft}        &  769 & 825 & 890 & 1002  \\
 {\rm HLPSStiff}       &  460 & 490 & 518 & 544  \\
 {\rm IU-FSU}          &  535 & 590 & 661 & ---  \\
 {\rm IU-FSUmax}       &  504 & 563 & 638 & ---  \\
\hline
\end{tabular}
\caption{Maximum initial frequency for neutron stars with masses of
$1.4$, $1.6$, $1.8$ and $2.0$ $M_{\odot}$ whose crust never fails
from slowing down (See Fig.~\ref{Fig2}). The maximum stellar masses
predicted by the IU-FSU and the IU-FSUmax models are
$1.94M_{\odot}$, hence the last column is left empty.
\label{Table1}}
\end{table}

Now we consider crust breaking for isolated stars that are born with
some initial frequency and are spinning down due to \textit{e.g.}
magnetic dipole radiation. As the star loses angular momentum, the
fluid in the core responds by reducing the equatorial bulge and it
becomes more spherical. Strain develops in the solid crust that may
eventually fail after losing support from the core, see Fig.
\ref{Fig1b} (Right). However, if the crust is strong enough, then
there is a certain maximum initial frequency below which the crust
will never break, even when the star has (nearly) stopped rotating.
In Fig.\,\ref{Fig2} we plot this maximum initial frequency as a
function of stellar mass for various EOS models, where we assume
$\sigma = 0.1$\,\cite{Horowitz:2009ya,Chugunov2010}. In
Table\,\ref{Table1} we provide predicted values of the maximum
initial frequency for several NS masses. The solution suggests that
the limiting spin frequency is proportional to the square root of
the breaking strain, $f_{\rm max} \propto \sqrt{\sigma}$, therefore
results presented in Table\,\ref{Table1} as well as Fig.\,\ref{Fig2}
can be easily scaled to other values of $\sigma$. In particular, the
crust of a canonical $1.4 M_{\odot}$ NS, assuming the HLPSStiff EOS
and a breaking strain of $\sigma = 0.1$, will never break if it is
born with $f \lesssim 460$ Hz or correspondingly with the spin
period of $P \gtrsim 2.17$ ms. This suggests that {\it the crust of
most isolated neutron stars may never have failed from slowing
down}.

Alternatively, if a NS is born with a large spin frequency above
this value, the crust could break after the star has spun down
significantly.  A very large initial spin frequency could occur, for
example, in a merger of two low mass NS that produces a stable very
rapidly rotating NS as a remnant.  If the crust then failed
asymmetrically, the star will have a nonzero $\epsilon$ and will
radiate gravitational waves.  Thus young NSs, produced in exotic
events with large initial spins, may be promising sources for
continuous GW searches.


\begin{table}[h]
\begin{tabular}{|l||c|c|c|c|c|}
 \hline
Model                  & $\sigma$ & $f_{\rm in}^{1.4}$ (Hz)             & $f_{\rm fin}^{1.4}$   (Hz)        & $f_{\rm in}^{1.8}$  (Hz) & $f_{\rm fin}^{1.8}$  (Hz)  \\
 \hline
 \hline
 {\rm HLPSStiff}       & 0.05 &  ~~~0 & ~326 & ~~35  & ~368  \\
                       & 0.10 &  ~136 & ~479 & ~236  & ~569  \\
 {\rm IU-FSU   }       & 0.05 &  ~349 & ~515 & ~909  & 1022  \\
                       & 0.10 &  ~781 & ~947 & 1875  & 1988  \\
 {\rm IU-FSUmax}       & 0.05 &  ~~35 & ~358 & ~374  & ~586 \\
                       & 0.10 &  ~232 & ~555 & ~854  & 1066  \\

\hline
\end{tabular}
\caption{Initial and final rotational frequencies of accreting
neutron stars whose crust breaks when the crust is fully replaced.
The results are presented for breaking strains of $\sigma = 0.05$
and $0.1$ and for two different stellar masses of $M = 1.4
M_{\odot}$, $1.8 M_{\odot}$. The crust of compact stars as predicted
by the HLPSSoft EOS never breaks for these values of $\sigma$ and
therefore results for this EOS are not listed.\label{Table2}}
\end{table}

We now consider NSs in accreting systems that increase their
rotational frequency via the transfer of angular momentum from the
companion. For example, we consider the case of recycled pulsars. We
consider a very simple model of how strain develops in the crust as
the crust is being replaced. We assume the crust starts from zero
strain at some initial frequency $f_{\rm in}$ such that the critical
strain is developed by the time the crust is fully replaced and the
star has been spun up to $f_{\rm fin}$.  Future work should explore
the stress in partially replaced crusts in more detail.

The average spin frequency derivative of an accreting pulsar can be estimated using some simple
accretion model\,\cite{Patruno:2012}:
\begin{equation}
\dot{f} = 2.3 \times 10^{-14} \sqrt{\xi} \dot{M}^{6/7}_{-10}
M^{3/7}_{1.4} B^{2/7}_8 R^{6/7}_{10} {\rm Hz\,s}^{-1} \ ,
\label{freqincrease}
\end{equation}
where $\xi \approx 0.3 - 1.0$ is a correction factor due to the
non-spherical geometry of accretion, $\dot{M}_{-10}$ is the mass
accretion rate in units of $10^{-10} M_{\odot} \,\rm{year}^{-1}$,
$R_{10}$ is the radius of the neutron star in units of 10 km, and
$B_{8}$ is the surface magnetic field of the neutron star in units
of $10^{8}$ Gs. For the sake of simplicity, we take $\xi = 1$,
$\dot{M}_{-10} =1$, and $B_8 = 1$. As neutron stars accrete
materials, their initial spin frequency changes by $\delta f \simeq
\dot{f} \Delta t$. The crust is fully replaced in a timescale of
$\Delta t \simeq M_{\rm crust}/\dot{M}$ years. During this time
period the spin frequency of the neutron star changes by $\delta f$,
and stresses develop in the newly formed crust. The largest strain
angle develops at the crust-core boundary near the
equator\,\cite{Franco:2000} and is therefore the first place where
the crust fails.
In Table~\ref{Table2} we
present predictions for the initial $f_{\rm in}$ and final $f_{\rm fin}$
rotational frequencies of neutron stars for several stellar
configurations and values of the breaking strain. The crust of
compact stars---as predicted by the HLPSSoft---do not develop large
strain angles and therefore require smaller values of the breaking
strain in order to fail. In other words, with $\sigma \simeq 0.1$ the crust of a very
compact star may never break. The effect of compactness is also
clearly visible when results are compared between 1.4 and 1.8
$M_{\odot}$ NSs whose radii are more or less constant. On the other hand, the
crust of large stars (HLPSStiff), as well as those with a thick
crust (IU-FSUmax) are more likely to fail even if the material
breaking strain is large.  In particular, for a canonical NS in the
IU-FSU model, we find that an increase in rotational frequency from
$f_{\rm in} = 349$ Hz to $f_{\rm fin} = 515$ Hz is needed before its
crust is replaced and a strain angle of as large as $\varepsilon =
\sigma = 0.05$ is developed.  Note that a larger value of the
breaking strain shifts these values to higher frequencies (See
Table~\ref{Table2}).

\begin{figure}[ht]
\smallskip
 \includegraphics[width=1.0\columnwidth]{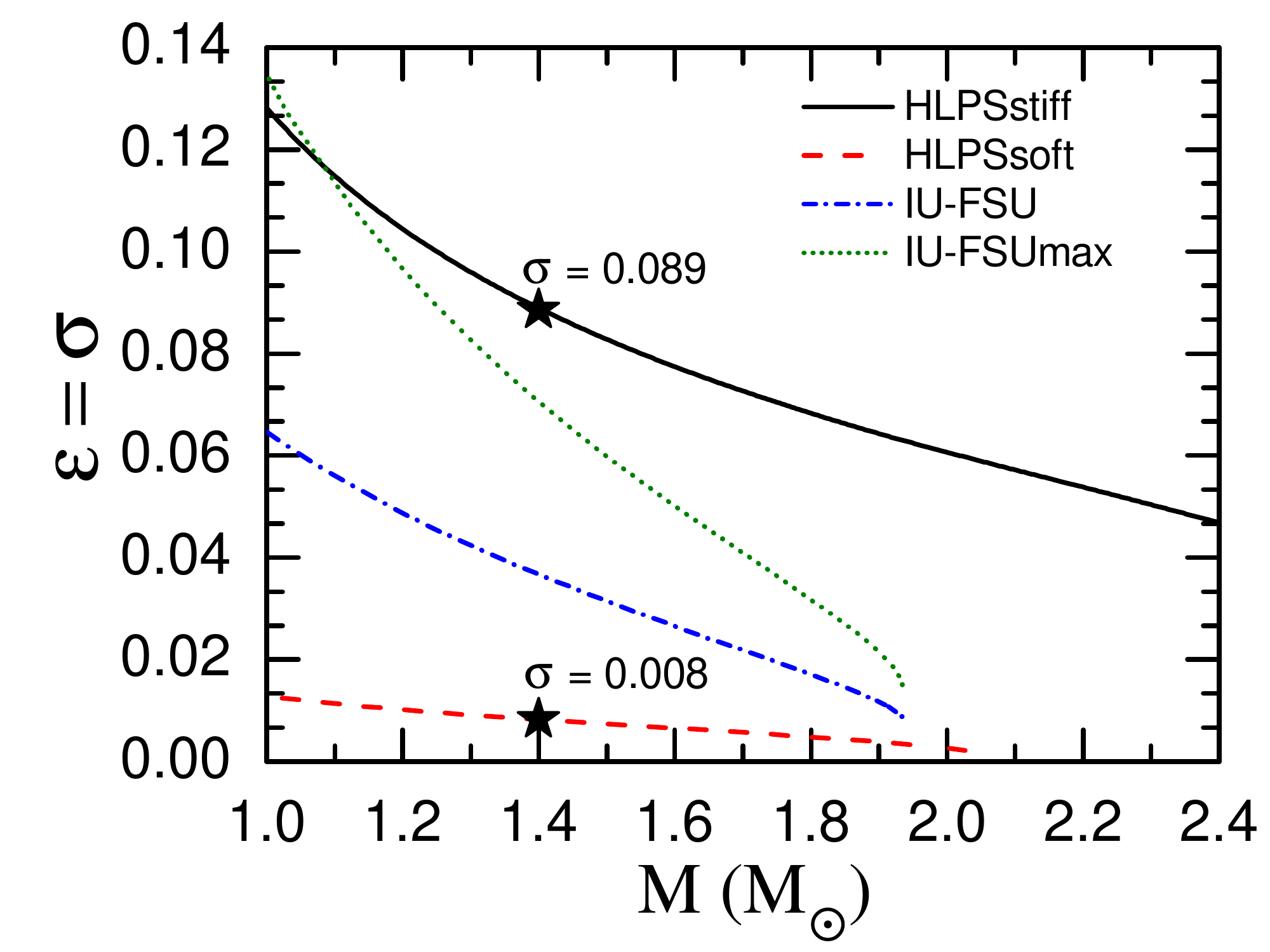}
 \caption{(Color online). The breaking strain $\sigma$ as a function of the neutron
 star mass that allows the crust to fail when the final rotational frequency $f_{\rm fin} = 716.36$ Hz is equal to the maximum
 observed frequency.}
 \label{Fig3}
\end{figure}
Instead of fixing the breaking strain, in Fig.\,\ref{Fig3} we now
plot strain angle at the equatorial crust-core boundary after the
crust has been fully replaced and the NS has reached the maximum
observed frequency of $f_{\rm fin} = 716.36$
Hz\,\cite{Hessels:2006ze}.  As discussed earlier, compact stars
require a relatively small breaking strain to fail. Whereas large
stars need a very strong crust to spin up to the observed
frequencies without breaking. In other words, for a fixed material
breaking strain and a given change in the rotational frequency, the
crust of large stars are more likely to fail than that of compact
stars. Similarly, neutron stars with a thick crust are also more
likely to break by the time the crust is fully replaced, mainly
because the total accretion torque is proportional to the mass of
the replaced crust (See also Fig.\ref{Fig1}). In particular, we find
that the required breaking strain for a $1.4 M_{\odot}$ neutron star
can be between $0.008 \lesssim \sigma \lesssim 0.089$ (See
Fig.\,\ref{Fig3}). This result is quite interesting, because it is
the same order as the breaking strain predicted by large scale
molecular dynamics simulations of crust
breaking\,\cite{Horowitz:2009ya}.

In summary, we have explored the impact of the change in rotational
frequency of NSs on the strong crust. We found that the crust of
most isolated NSs may never fail. On the other hand, the crust of
NSs in accreting systems is found to fail at rotational frequencies
in agreement with observations. When the crust fails, a nonzero
$\epsilon$ is likely to be produced and the star will start to
radiate GW, see Fig. \ref{Fig1b} (Left). Torque from the GW
radiation is a stiff function of the rotational frequency.  If the
initial GW torque is larger than the accretion torque, the star will
spin down slightly. Alternatively, if the initial GW torque is
smaller than the accretion torque, the star will continue to spin up
slightly. In both cases the star is expected to reach equilibrium
where the GW torque balances the accretion torque.  This will leave
the accreting star a potential promising source for present and
future searches for continuous GW.

\begin{acknowledgments}
 This material is based upon work supported by the U.S. Department
 of Energy Office of Science, Office of Nuclear Physics under Awards
 DE-FG02-87ER40365 (Indiana University) and DE-SC0018083 (NUCLEI SciDAC-4 Collaboration).
\end{acknowledgments}

\bibliography{ReferencesFJF}

\end{document}